\documentclass[12pt]{article}
\usepackage{graphicx}
\usepackage{amsfonts}
\usepackage{amssymb}
\usepackage{color}
\input{psfig}

\begin{document}

\def\journal{\topmargin .3in    \oddsidemargin .5in
        \headheight 0pt \headsep 0pt
        \textwidth 5.625in 
        \textheight 8.25in 
        \marginparwidth 1.5in
        \parindent 2em
        \parskip .5ex plus .1ex         \jot = 1.5ex}

\def\hybrid{\topmargin -20pt    \oddsidemargin 0pt
        \headheight 0pt \headsep 0pt
        \textwidth 6.25in
        \textheight 9.00in       
        \marginparwidth .875in
        \parskip 5pt plus 1pt   \jot = 1.5ex}


\journal

\catcode`\@=11

\def\marginnote#1{}
%
\newcount\hour
\newcount\minute
\newtoks\amorpm
\hour=\time\divide\hour by60
\minute=\time{\multiply\hour by60 \global\advance\minute by-\hour}
\edef\standardtime{{\ifnum\hour<12 \global\amorpm={am}%
        \else\global\amorpm={pm}\advance\hour by-12 \fi
        \ifnum\hour=0 \hour=12 \fi
        \number\hour:\ifnum\minute<10 0\fi\number\minute\the\amorpm}}
\edef\militarytime{\number\hour:\ifnum\minute<10 0\fi\number\minute}

\def\draftlabel#1{{\@bsphack\if@filesw {\let\thepage\relax
   \xdef\@gtempa{\write\@auxout{\string
      \newlabel{#1}{{\@currentlabel}{\thepage}}}}}\@gtempa
   \if@nobreak \ifvmode\nobreak\fi\fi\fi\@esphack}
        \gdef\@eqnlabel{#1}}
\def\@eqnlabel{}
\def\@vacuum{}
\def\draftmarginnote#1{\marginpar{\raggedright\scriptsize\tt#1}}

\def\draft{\oddsidemargin -.5truein
        \def\@oddfoot{\sl preliminary draft \hfil
        \rm\thepage\hfil\sl\today\quad\militarytime}
        \let\@evenfoot\@oddfoot \overfullrule 3pt
        \let\label=\draftlabel
        \let\marginnote=\draftmarginnote
   \def\@eqnnum{(\theequation)\rlap{\kern\marginparsep\tt\@eqnlabel}%
\global\let\@eqnlabel\@vacuum}  }


\def\preprint{\twocolumn\sloppy\flushbottom\parindent 2em
        \leftmargini 2em\leftmarginv .5em\leftmarginvi .5em
        \oddsidemargin -.5in    \evensidemargin -.5in
        \columnsep .4in \footheight 0pt
        \textwidth 10.in        \topmargin  -.4in
        \headheight 12pt \topskip .4in
        \textheight 6.9in \footskip 0pt
        \def\@oddhead{\thepage\hfil\addtocounter{page}{1}\thepage}
        \let\@evenhead\@oddhead \def\@oddfoot{} \def\@evenfoot{} }

\def\proceedings{\pagestyle{empty}
        \oddsidemargin .26in \evensidemargin .26in
        \topmargin .27in        \textwidth 145mm
        \parindent 12mm \textheight 225mm
        \headheight 0pt \headsep 0pt
        \footskip 0pt   \footheight 0pt}


\def\numberbysection{\@addtoreset{equation}{section}
        \def\theequation{\thesection.\arabic{equation}}}

\def\underline#1{\relax\ifmmode\@@underline#1\else
        $\@@underline{\hbox{#1}}$\relax\fi}

\newcommand{\be}[1]{\begin{equation}\label{#1}} 
\newcommand{\ee}{\end{equation}}
\newcommand{\bea}[1]{\begin{eqnarray}\label{#1}} 
\newcommand{\eea}{\end{eqnarray}}

\newcommand{\ttil}{\tilde{t}}
\newcommand{\pt}{\tilde{\phi}}
\newcommand{\thetat}{\tilde{\theta}}

\newcommand{\apt}{\tilde{\alpha}'}
\newcommand{\srt}{\sqrt{2}}
\newcommand{\A}{{\cal A}}
\newcommand{\cA}{{\cal A}}\newcommand{\cB}{{\cal B}}
\newcommand{\cC}{{\cal C}}\newcommand{\cD}{{\cal D}}
\newcommand{\cE}{{\cal E}}\newcommand{\cF}{{\cal F}}
\newcommand{\cG}{{\cal G}}\newcommand{\cH}{{\cal H}}
\newcommand{\cI}{{\cal I}}\newcommand{\cJ}{{\cal J}}
\newcommand{\cK}{{\cal K}}\newcommand{\cL}{{\cal L}}
\newcommand{\cM}{{\cal M}}\newcommand{\cN}{{\cal N}}
\newcommand{\cO}{{\cal O}}\newcommand{\cP}{{\cal P}}
\newcommand{\cQ}{{\cal Q}}\newcommand{\cR}{{\cal R}}
\newcommand{\cS}{{\cal S}}\newcommand{\cT}{{\cal T}}
\newcommand{\cU}{{\cal U}}\newcommand{\cV}{{\cal V}}
\newcommand{\cW}{{\cal W}}\newcommand{\cX}{{\cal X}}
\newcommand{\cY}{{\cal Y}}\newcommand{\cZ}{{\cal Z}}

\newcommand{\ra}{\rightarrow} \newcommand{\wg}{\wedge}
\newcommand{\ap}{\alpha'} \newcommand{\Tr}{\mbox{Tr}}

\def\titlepage{\@restonecolfalse\if@twocolumn\@restonecoltrue\onecolumn
     \else \newpage \fi \thispagestyle{empty}\c@page\z@
        \def\thefootnote{\fnsymbol{footnote}} }

\def\endtitlepage{\if@restonecol\twocolumn \else \newpage \fi
        \def\thefootnote{\arabic{footnote}}
        \setcounter{footnote}{0}}  

\catcode`@=12
\relax

\begin{titlepage}

\hfill
\vbox{\halign{#\hfil          \cr
             UTTG-15-02       \cr
             CERN-TH/2002-367 \cr
             hep-th/0212147   \cr}} 
\vspace*{15mm}
\begin{center}
{\Large {\bf  Large Spin Strings in $AdS_3$}\\} 
\vspace*{10mm}
Amit Loewy$^a$
\footnote{e-mail: {\tt loewy@physics.utexas.edu}}
 and Yaron Oz$^{b,c}$
\footnote{e-mail: {\tt yaronoz@post.tau.ac.il, Yaron.Oz@cern.ch}}\\ 
\vspace*{10mm} 

{\it {$^{a}$ Department of Physics, \\
University of Texas, Austin, TX 78712}}\\

\vspace*{5mm}
{\it {$^{b}$ School of Physics and Astronomy,\\
Raymond and Beverly Sackler Faculty of Exact Sciences\\
Tel-Aviv University, Ramat-Aviv 69978, Israel}}\\ 

\vspace*{5mm}
{\it {$^{c}$Theory Division, CERN \\
CH-1211 Geneva  23, Switzerland}}\\

\end{center}

\vspace*{5mm}
\begin{abstract}
We consider strings with large spin in $AdS_3\times S^3\times 
{\cal M}$ with NS-NS background. 
We construct the string configurations as solutions
of $SL(2,R)$ WZW theory.
We compute the relation between the space-time energy and spin, and show 
that the anomalous correction is constant, and not logarithmic in
the spin. This is in contrast to the S-dual background with
R-R charge where the anomalous correction is logarithmic.
\end{abstract}


\end{titlepage}
\newpage

\section{Introduction}

Examples where gauge theory perturbative 
computations can be extrapolated to strong coupling and compared to
classical supergravity results are rare when the
calculated quantities are not protected by supersymmetry.  
An important example has been studied by Gubser, Klebanov and Polyakov \cite{GKP}.
They showed that the energy of 
a spinning string  with spin $S$ in $AdS_5$ 
in the limit that its size is much larger than
the $AdS_5$ radius is
\be{GKPr}
E = S + \frac{\sqrt{\lambda}}{\pi} \ln (S/\sqrt{\lambda}) \ . \ee
In global coordinates the energy is identified with the
dimension of the dual operator.
GKP proposed that this string configuration is dual to a twist 2
operator in $\cN=4$ SYM such as 
\be{t2o}
\cO_{(\mu_1 \cdots \mu_n)} \ = \ \Tr \  \phi^* \cD_{(\mu_1} \cdots
\cD_{\mu_n)} \phi \ . \ee
In perturbative gauge theory this operator has an anomalous dimension that grows
as $\ln n$, which is in agreement with the strong coupling result
derived from supergravity. Quantum corrections to the energy of
the spinning string were analyzed  in \cite{FT,Tseytlin} with
no $\ln^2 S$ corrections found.
One therefore  expects that the leading term in the anomalous dimension
to all orders in perturbation theory (and probably non-perturbatively
as well) is logarithmic, i.e.
\be{GKPr}
E = S + f(\lambda) \ln (S/\sqrt{\lambda}) \ . \ee
Following the work of GKP many other string and membrane
configurations were studied \cite{Russo} -- \cite{RV}. For earlier
works on semi-classical string solutions see \cite{NS}.

In this paper we will study spinning strings in $AdS_3 \times S^3
\times \cM$ with NS-NS 2-form background. 
This is of interest, since string theory on
this background can be exactly solved in terms of the
$SL(2,R)$ WZW model \cite{Giveon:1998ns,MO}. 
We will see that the ``planetoid''
spinning string solution of $AdS_5$ is not a solution
in $AdS_3$ because
of the NS-NS 2-form. In section 2 we will find the explicit form of 
the classical spinning string configuration, and
calculate the energy--spin relation. We will see that the leading
behavior is the same as in $AdS_5$, but the anomalous correction is
constant rather than logarithmic. 
This is in contrast with the S-dual background with
R-R flux (for a review see \cite{Aharony:1999ti}),
where the anomalous correction is logarithmic.
Indeed it confirms the fact that the near horizon limits of
D1-D5 and F1-NS5 systems lead to very different theories. 
In section 3 we will analyze the
conditions under which the classical string configuration is valid as
a quantum state. In section 4 we consider more
general string configurations that rotate on the $S^3$ part of
the metric as well. In the last section we discuss the results.

\section{Classical spinning strings in $AdS_3$}

String theory on $AdS_3 \times S^3 \times \cM$ has many classical
solutions. We will be interested in a class of classical
solutions that correspond to spinning closed strings
\cite{FHV,VE,KM}. 
Similar string configurations were also analyzed in \cite{LL} 
in the context of open strings. 

The NS-NS $AdS_3$ background, in global coordinates, is given by
\be{ads3}
ds^2 = R^2 \left( \ -(1+r^2)dt^2 + \frac{dr^2}{1+r^2} + 
r^2 d \phi^2 \ \right) \ , \ee
$$ 
\mbox{with} \qquad 
B_{t \phi} = R^2r^2 \ . 
$$
We will consider 
the following ansatz for a time dependent embedding of a closed
string in $AdS_3$
\be{ansatz} 
t= c_1 \tau + \ttil \ (\sigma) \ , 
\ee
$$
\phi=c_2 \tau +\pt \ (\sigma) \ , 
$$
$$
r= r \ (\sigma) \ . 
$$  
$c_1$ and $c_2$ are constants, and $c_1$
is assumed to be positive to insure forward propagation in time.
$\tau$ and $\sigma$ denote the world-sheet coordinates 
and $\sigma \simeq \sigma+ 2\pi$.

The GKP ansatz is the one in which $\ttil = \pt = 0$ . At
first sight this seems to be a solution of the classical
equations of motion. The
energy-spin relation for such a configuration will be as in
\cite{GKP}. For
strings larger than the AdS radius,
it takes the form (\ref{GKPr}), with $\sqrt{\lambda}=R^2/\ap \ $.
However, we will see that while this configuration solves
the constraints $T_{\pm
\pm}=0$, the $SL(2,R)$ Kac-Moody
currents associated with it
are not holomorphic. Note, however, that for a fundamental spinning
string in a R-R $AdS_3$ background, given by the near-horizon limit
of the D1-D5 system, the simpler ansatz is perfectly valid. 

The world-sheet action for a closed string embedding of the form
(\ref{ansatz}) reads
\be{action}
S = \frac{R^2}{2\pi\ap} \int d \tau d \sigma \ \Big[ 
-(1+r^2)\left(-c_1^2 + \left(\frac{d\ttil}{d\sigma}\right)^2 \ \right)+
\frac{1}{1+r^2}\left(\frac{dr}{d\sigma}\right)^2 + 
\ee
$$
+ \ r^2\left(-c_2^2+ \left(\frac{d\pt}{d\sigma}\right)^2 \ \right) 
-2 r^2 \left(c_1 \left(\frac{d\pt}{d\sigma}\right) - 
c_2 \left(\frac{d\ttil}{d\sigma}\right)\right) \Big] \ . 
$$
From (\ref{action}) we get two conservation laws 
\be{conservationlaws}
\frac{d\ttil}{d\sigma} = \frac{c_2r^2-k_1}{1+r^2} \ , 
\ee
$$   
\frac{d\pt}{d\sigma} = \frac{c_1r^2-k_2}{r^2} \ ,
$$
where $k_1$ and $k_2$ are integration constants.
The constraint on the energy-momentum tensor of the system is
\be{constraints}
T_{\pm \pm}=T_{\sigma \sigma}+T_{\tau \tau} \pm 2T_{\tau \sigma}=0 \ .
\ee
In (\ref{constraints}) we assumed that there are no other contributions
to the energy-momentum tensor from the internal CFT on $S^3 \times \cM$.
It is, however, easy to generalize this calculation to include
contributions from the internal CFT. In such a case the constraint is on
the sum of energy-momentum tensors 
\be{}T^{AdS}_{\pm \pm} + T^{S^3 \times \cM}_{\pm \pm} = 0 \ . \ee 
We will discuss this generalization 
in section 4.
From the requirement that $T_{\sigma \tau}=0$ we get that
$c_1k_1=c_2k_2 \ .$ To simplify the notation we define
\be{} \alpha=k_1^2-k_2^2 \ ,\ee
$$\beta=c_1^2+2c_1k_2 \ .$$ 
With these definitions the energy-momentum constraint reduces to 
\be{dr}
\left(\frac{dr}{d\sigma}\right)^2 = \frac{\alpha \beta}{k_2^2r^2}
\left( r^2-\frac{k_2^2}{\beta} \right) \
\left(\frac{k_2^2}{\alpha}-r^2 \right) \ .  
\ee
The solution to this equation is given by
\be{r(s)} 
r^2(\sigma) = \frac{k_2^2}{2\alpha \beta} \left(
(\alpha+\beta) - (\alpha -\beta) \sin \left[\frac{\sqrt{4\alpha \beta}}{k_2}
\sigma \right] \right) \ .
\ee
For a closed string we need to impose $r(2\pi)=r(0)$ which leads
to the quantization of $n$.
\be{n} 
n = \frac{\sqrt{4\alpha \beta}}{k_2} = \mbox{integer} \ .
\ee
The angular coordinate can be deduced from (\ref{ansatz}) and
(\ref{conservationlaws}) to be
\be{phi} 
\phi(\sigma , \tau) \ = \ \frac{c_1k_1}{k_2} \  \tau +
c_1 \sigma - \arctan \left[ \frac{(\alpha+\beta)\tan
\left[\frac{n}{2}\sigma
\right]- (\alpha-\beta)}{\sqrt{4\alpha \beta}} \right] \ . 
\ee
Note that (\ref{phi}) is discontinuous whenever $\tan
[\frac{n}{2}\sigma]$ diverges. In order to
get a continuous function we should add $\pi$ after each discontinuous
point ($n \pi$ altogether). 
In order to ensure the periodicity $\phi(2\pi,\tau)=\phi(0,\tau)$ mod 
$2\pi$, we should have either (i) $c_1$ an integer and $n$ even, or (ii)
$c_1$ half integer and $n$ odd. A few examples are plotted in figure
1 for $\tau=0$.
\newpage 
\begin{figure}
\leavevmode
\centerline{
\psfig{figure=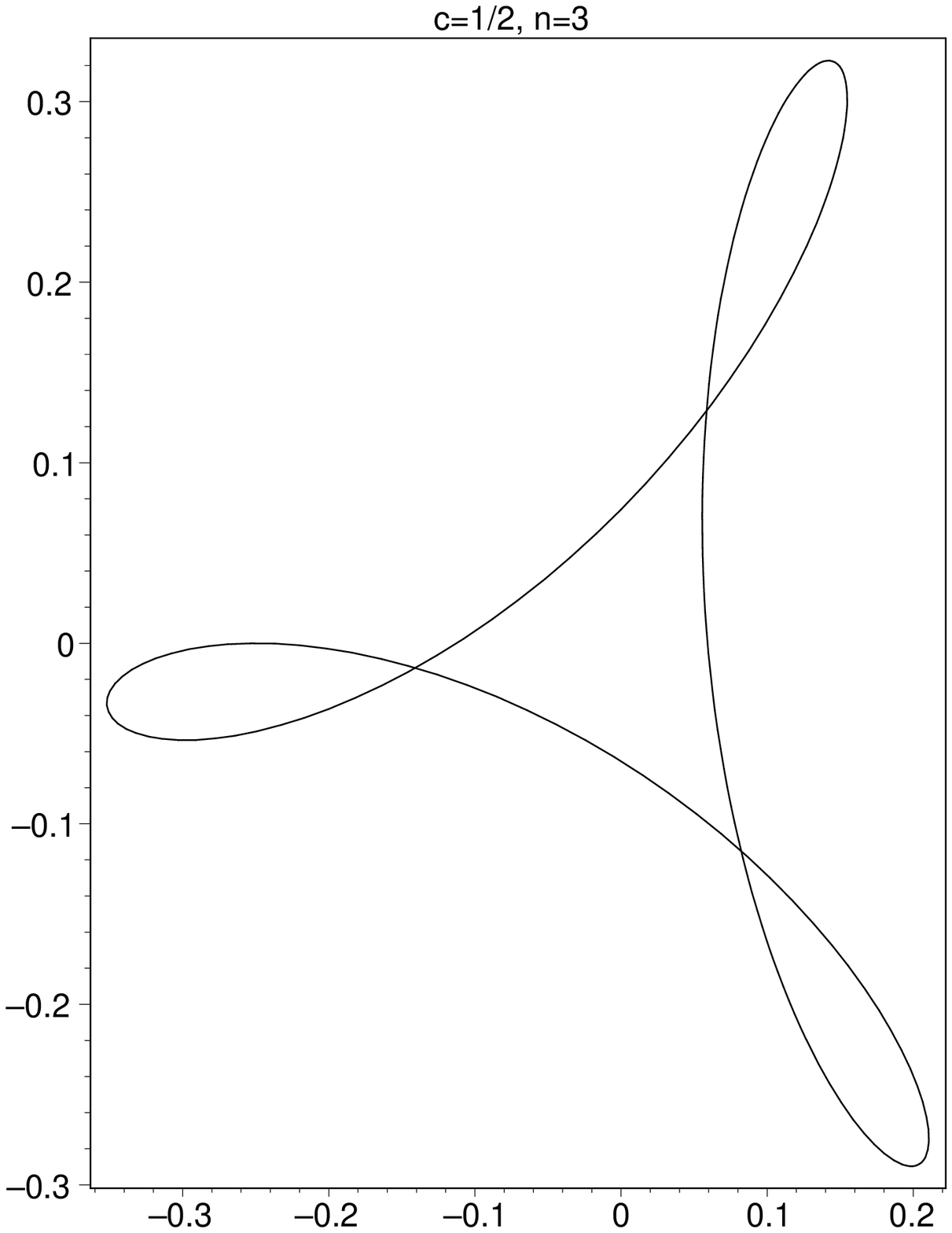,width=7cm,height=7cm,clip=}
\psfig{figure=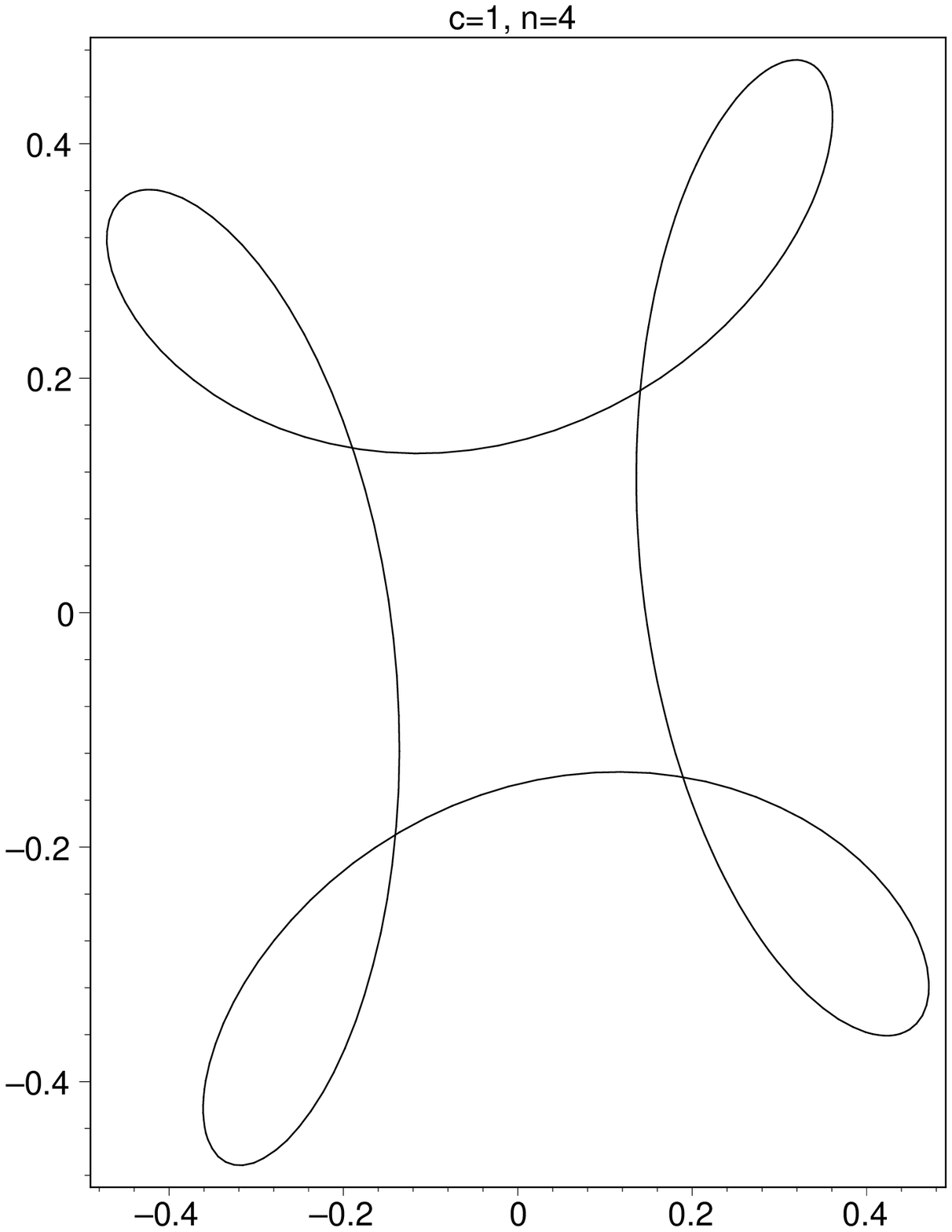,width=7cm,height=7cm,clip=}
}
\caption{Parametric polar plots  
depicting the string at $\tau=0$ for
(i) $n=3$, $c_1=1/2$. (ii) $n=4$, $c_1=1$. The radial coordinate is
$r(\sigma)$ and the angular coordinate is $\phi(\sigma)$. Note that
these plots show the string at constant world-sheet time, which is not the same
as target space time because of (\ref{ansatz}).}
\end{figure}

The time coordinate is given by
\be{t} 
t(\sigma, \tau) \ = \ 
c_1 \tau \ + \ c_2 \sigma \ + 
\ee
$$
- \ \frac{(c_2+k_1)\sqrt{4 \alpha \beta}}
{n \sqrt{(k_2^2+\alpha)(k_2^2+\beta)}} 
\arctan 
\left[ \frac{\left(1+\frac{k_2^2(\alpha +\beta)}{2\alpha \beta}
\right) \tan \left[\frac{n}{2}\sigma \right] - \frac{k_2^2(\alpha-\beta)}
{2\alpha \beta}}{\sqrt{\frac{(k_2^2+\alpha)(k_2^2+\beta)}
{\alpha \beta}}} \right] \ .
$$
The periodic boundary condition $t(2\pi,\tau)=t(0,\tau)$ is obeyed if $k_2$
satisfies (again note that $\arctan[...]$ gets shifted by $n\pi$)
\be{c_2} 
c_2 \ = \ \frac{(c_2+k_1)\sqrt{\alpha \beta}}
{\sqrt{(k_2^2+\alpha)(k_2^2+\beta)}}  \ . 
\ee
This can be solved using the definitions of $\alpha$ and $\beta$,  and
gives the allowed value for $k_2$  
\be{k_22} 
k_2 = \frac{2c_1^3}{n^2-4c_1^2} \ . 
\ee
Note that the values of $n$ and $c_1$ should be restricted to $n>2c_1
\ $. At the point $n=2c_1$ the size of the string diverges. We will
analyze this limit below. For $n<2c_1$ we have
the unphysical result $r^2(\sigma)<0$.
Substituting $k_2$
back in equations (\ref{r(s)}), (\ref{phi}) and (\ref{t}) we get
\be{rs1}
r^2(\sigma) \ = \ \frac{2}{n^2} \left(
\frac{c_1^2(n^2+c_1^2)}{n^2-4c_1^2} +
\frac{c_1^2(n^2-c_1^2)}{n^2-4c_1^2} \sin [ n \sigma ] \right) \ ,
\ee  

\be{phiphi}
\phi(\sigma , \tau) = \frac{n}{2} \  \tau +
c_1 \sigma - \arctan \Big[ \ 
\frac{n^2+c_1^2}{2c_1n} \tan \left[\frac{n}{2}\sigma \right]
+ \frac{n^2-c_1^2}{2c_1n} \Big] \  , \ee

\be{tt} 
t(\sigma, \tau) = c_1 \tau + \frac{n}{2} \sigma - \arctan 
\Big[ \ 
\frac{n^4-2c_1^2n^2+2c_1^4}{n^2(n^2-2c_1^2)} \tan 
\left[\frac{n}{2}\sigma \right] + \frac{2c_1^2(n^2-c_1^2)}{n^2(n^2-2c_1^2)}
\ \Big] .
\ee

The space-time energy $E$ and spin $S$ of the
system are given by
\be{concha}
E = \frac{R^2}{2\pi\ap} \int_{0}^{2\pi} d\sigma \ \frac{\delta L}{\delta
\dot{t}} \ = \ \frac{R^2}{2 \pi \ap} \int_{0}^{2\pi} d\sigma \ 
\left( c_1 (1+r^2) - r^2 \frac{d\phi}{d\sigma} \right) \ , 
\ee
$$
S = \frac{R^2}{2\pi \ap} \int_{0}^{2\pi} d\sigma \ \frac{\delta L}
{\delta \dot{\phi}} \ = \ \frac{R^2}{2\pi \ap} \int_{0}^{2\pi} d\sigma \
\left(-c_2 r^2 + r^2 \frac{dt}{d\sigma} \right) \ .
$$
Using (\ref{conservationlaws}), the integrals can be calculated
explicitly, and we can express  
$E$ and $S$ in terms of $c_1$ and
$n$ as
\be{concha1}
E = \frac{R^2(c_1+k_2)}{\ap} \ = \ \frac{R^2 c_1}{\ap} \
\frac{n^2-2c_1^2}{n^2-4c_1^2} \ ,
\ee
$$
S \ =  \ \frac{R^2 k_1}{\ap} \ = \ \frac{R^2}{\ap}
\ \frac{c_1^2 n}{n^2-4c_1^2} \ .
$$
It is straightforward to solve this system and get
the energy-spin relation 
\be{E(s)}
E(S,n) \ = \ \frac{R^2}{\ap} \frac{n^2+2n^2 \ap S/R^2}{n^3}
\sqrt{\frac{n^2 \ap S /R^2}{n+4 \ap S /R^2}} \ . \ee
 
Let us examine this relation in the limits where the string size is
much smaller or much larger than the $AdS_3$ radius.
These two limits will correspond to the $2c_1 \ll n$ and the $2c_1 \ra
n$  limits, respectively.
When $2c_1 \ll n$ the string is mostly concentrated near the origin
of $AdS_3$, as can be seen from taking the limit of (\ref{rs1})
\be{limit1r(s)}
r^2(\sigma) \ra \frac{2 c_1^2}{n^2} (1+\sin [n \sigma]) \ .
\ee
In other words, the string size is much less than the $AdS_3$ radius.
In this limit we get the flat space
relation between the energy of the string and its spin
\be{nrainfty} 
S \ = \ \frac{\ap E^2}{R^2 n} \ .
\ee
On the other hand, if we take $\epsilon = n^2-4c_1^2 \ra 0$ the
asymptotic form of (\ref{rs1}) is
\be{limit2r(s)}
r^2(\sigma) \ra \frac{n^2}{8 \epsilon} (5+3 \sin [n \sigma]) \ .
\ee
In this limit the size of the string is much larger than the $AdS_3$ radius,
and the leading relation between $E$ and $S$ is given by
$E=S \ $.
This linear behavior was also found for open rotating strings in
\cite{LL}. Using (\ref{concha1}) we can actually calculate all the classical
corrections to the leading linear behavior
\be{}
E \ = \ S + \frac{3}{8} \frac{R^2n}{\ap} - \frac{10}{256}
\frac{R^4n^2}{\ap^2 S} + {\cal O}\left( \frac{R^6n^3}{\ap^3S^2}
\right) \ .
\ee
The first correction is a constant. Note in
comparison that in $AdS_5$ or $AdS_3$ with R-R background the
leading correction is $\ln S$.

\section{Quantum spinning strings}

The spectrum of strings on  $AdS_3$ with NS-NS background
was analyzed in \cite{MO}. In this section we will use a similar
analysis in order to
identify the $SL(2,R)$ representation 
corresponding to the
spinning string solution. We will investigate when
is this state expected to be
part of the physical spectrum of strings
on $AdS_3$. 

In order to identify the corresponding representation of $SL(2,R)$, 
we switch to light-cone coordinates, and calculate
the $SL(2,R)$ currents associated with the spinning string solution.
The solution of the previous section
satisfies $T_{\pm \pm}^{AdS}=0$, which implies that
the Casimir of $SL(2,R)$, $C_2=-j(j-1)=0$. We therefore should look
for a representation with $j=1$.  

We define $k=R^2 / \ap$. 
We also define new target space coordinates
\be{uv}
u=\frac{1}{2}(t+ \phi ) \ , \qquad v=\frac{1}{2}(t-\phi ) \ .
\ee
The light-cone coordinates on the world-sheet are: 
$x^{\pm}=\tau \pm \sigma \ $.
Using the conservation laws (\ref{conservationlaws}) we evaluate the 
right and left $SL(2,R)$ currents. In the following we shall only need
the explicit form of $J^3$  
\be{rlc}
J^3_R = k \left( \partial_+u+(1+2r^2) \partial_+ v \right) 
=k ( c_1+k_2-k_1) \ , 
\ee
$$
J_L^3= k \left( \partial_-v + (1+2r^2) \partial_- u \right) 
=k (c_1+k_2+k_1) \ .   
$$
The other two components, $J^{\pm}$, can
be determined in the same way.

Note that had we used the ansatz with $\ttil = \pt =0$ the currents
would have been
\be{}
J^3_R = k \left( (c_1+c_2)+(1+2r^2(\sigma)) (c_1-c_2) \right)\ , 
\ee
$$
J_L^3= k \left( (c_1-c_2) + (1+2r^2(\sigma)) (c_1+c_2) \right)\ .  
$$
By taking $c_1=\pm c_2$ we can the make left/right current
holomorphic/anti-holomorphic, but not both. In WZW theory the holomorphy of
the currents is equivalent to the equations of motion, thus such a
configuration does not solve the field equations of the
$SL(2,R)$ WZW model.
 
The zero modes of the currents (\ref{rlc}) are 
\be{zm}
J_0^3=\int_0^{2\pi} \frac{dx^+}{2\pi} J_R^3 \ , \qquad 
{\bar J}_0^3=\int_0^{2\pi} \frac{dx^-}{2\pi} J_L^3 \ .
\ee
We denote by $m$ and ${\bar m}$ the eigenvalues of $J_0^3$ and
${\bar J}_0^3$ which are given by 
\be{mmbar}
m=k \frac{n^2c_1-2c_1^3-nc_1^2}{n^2-4c_1^2} \ , \qquad 
{\bar m} = k \frac{n^2c_1-2c_1^3+nc_1^2}{n^2-4c_1^2} \ .
\ee
Because the currents (\ref{rlc}) are constant, the solution 
has only a zero mode and no other higher modes, since 
$J^3_{\nu} \sim \int J^3_L e^{i \nu x^+} dx^+$
vanishes unless $\nu=0$.
For this solution to be a valid quantum state it must be in a unitary
representation of $SL(2,R)$. The full quantum spectrum will be
associated with the corresponding representation of the affine
${SL(2,R)}$ subject to the physical state constraint.
 
Recall that the unitary representations of the $SL(2,R)$ Lie algebra are:

\begin{itemize}

\item Discrete representations: 
$\cD_j^+$ and $\cD_j^-$ for real $j$, for which
$m=j,j \pm 1, j \pm 2, \cdots$.

\item Continuous representations: 
$\cC_j^a$ ($0 \le a <1$) for $j=1/2+is$ ($s$ is real); 
or for $1/2 < j < 1$ and $j-1/2 < |a-1/2|$, for which 
$m=a, a \pm 1, a \pm 2, \cdots$   

\item Identity representation: $j=0$ .

\end{itemize}

Our solution has $j=1$ so it cannot correspond to a unitary continuous 
representation. It can correspond to one of the discrete representations but 
only if $m$ and $\bar{m}$ are integers. If they are not integers they can 
still belong to a non-unitary continuous representation, but we will
not
consider them as
physical quantum states.
It is clear that to have integer $m$ and $\bar{m}$, $k$ must be
rational. Let us examine the spectrum for $k=1$ as an example. 
There do not seem to be solutions with 
integer $m$ and $\bar{m}$ for odd $n$ in this case. 
For even $n$ there are solutions. 
The first one is $n=24, c_1=6$. 
In this case the string state corresponds to 
the WZW state 
$$|j=1,m=5 \rangle \times |\bar{j}=1, \bar{m}=9 \rangle \ .$$
For higher values of $k$ there will be other solutions. 
It is not clear though whether all integers
can be generated by (\ref{mmbar}) for a given value of $k$.

It was argued in \cite{MO} that 
using spectral flow one can generate new classical
solutions
\be{sf} t(\tau,\sigma) \ra t(\tau,\sigma) + \omega \tau \ , \ee
$$  \phi(\tau,\sigma) \ra \phi(\tau,\sigma) +\omega \sigma \ ,$$
where $\omega$ is an integer.
These new solutions were shown to be important in understanding the
nature of long string states in $AdS_3$.
The $SL(2,R)$ currents, and the energy-momentum tensor are modified by the
spectral flow to  
\be{} J^3_0 \ \ra \ J^3_0 + \frac{k \omega}{2} \ , \ee
$$T_{++} \ra T_{++} - \omega J^3_0 - \frac{k}{4} \omega^2 \ ,$$
and the same for $\bar{J}_0^3$, and $T_{--}$. The spin
of the flowed state is the same as the un-flowed state, since it
is $S=m-\bar{m}$. The energy, however, is shifted by $k\omega$.
Since the energy tensor is modified by the spectral flow one must impose a
new physical state condition
\be{} T_{++}^{(\omega)} = T_{++} - \omega J^3_0 - \frac{k}{4} \omega^2 =
0 \ , \ee
and the same for $T^{(\omega)}_{--}$.
Clearly, since $T_{\pm \pm}=0$ for the original solutions, and $J_0^3
\neq \bar{J}_0^3$, there cannot be a solution to the physical state
condition other than $\omega=0$. In fact, this will also be true even
if there are contributions from the CFT on $S^3$, as long as
$T_{++}^{S^3}=T_{--}^{S^3}$. 

We now describe the standard way in which 
the quantum spectrum of strings in
$AdS_3 \times S^3 \times \cM$ is built around
the classical spinning string solution we presented.   
We consider the state $|\psi_0 \rangle = 
|j,\bar{j}, m, \bar{m}, h, \bar{h} \rangle$ as the ground
state, where $(h, \bar{h})$ are the conformal
weights of some state of the internal CFT on $S^3 \times
\cM$. In our case $h=\bar{h}=0$. 
The excited states  
are constructed by applying the operators
$J^a_{-\nu}$, with $\sum \nu_i=N$ and  $L_{-\mu_i}$ with 
$\sum \mu_i = M$ (and the same for the anti-holomorphic sector)
\be{es}
|\psi \rangle \ = \  \prod L_{-\mu_{i}} \prod 
\bar{L}_{-\bar{\mu}_{i}}\prod J^{a_i}_{-\nu_{i}} 
\prod \bar{J}^{a_i}_{-\bar{\nu}_{i}} |\psi_0 \rangle \ , \ee
and imposing 
the physical state conditions \cite{GW} 
\be{pyscon}
(L_0^{AdS_3}-1+N+M+h)|\psi \rangle \ = \
\left(-\frac{j(j-1)}{k-2}-1+N+M+h \right) |\psi \rangle \ = \ 0 \ , \ee
$$J_{\nu}|\psi \rangle = L_{\nu} |\psi \rangle = \ 0 \ , \qquad \nu >0\ , $$
and a similar constraint for the anti-holomorphic part. 
For the state to be invariant under arbitrary translation of the
world-sheet coordinate $\sigma$, one must also impose a level matching
condition
\be{} (L_0^{Total}-\bar{L}_0^{Total}) |\psi \rangle = 0 \ , \ee
which can also be stated as 
\be{} -\frac{j(j-1)}{k-2}+N+M+h \ = \
-\frac{\bar{j}(\bar{j}-1)}{k-2}+\bar{N}+\bar{M}+\bar{h}\ . \ee

\section{Adding momentum on $S^3$}

In this section we would like to generalize the discussion of spinning
strings in $AdS_3$ to include rotation in the $S^3$ part of the
metric. The simplest ansatz is to assume that the string is
point-like on $S^3$. The $S^3$ background is given by 
\be{su2b} ds^2 = R^2 ( \cos^2 x d\theta^2 - \sin^2 x d \thetat^2 - dx^2)
, \qquad B_{\theta \thetat} = 2 R^2 \cos^2 x \ , \ee
with the angular variables taken to be periodic 
\be{} \theta \sim \theta + 2\pi \ , \qquad \thetat \sim \thetat + 2 \pi \ .\ee 
Consider a string embedding in $S^3$ of the form $\theta=\omega \tau$,
and $\thetat=x=0$. 
The contributions to $T^{S^3}_{++}$ equals that for $T^{S^3}_{--}$
since the string embedding is $\sigma$ independent 
\be{} T_{\pm \pm}^{S^3} = R^2 \omega^2 \ .\ee
The physical state condition reads 
\be{}T^{AdS}_{\pm \pm}+ R^2\omega^2=0 \ . \ee 
This modifies (\ref{dr}) to be 
\be{drh}
\left(\frac{dr}{d\sigma}\right)^2 = \frac{\alpha \beta}{k_2^2r^2}
\left( r^2-\frac{k_2^2}{\beta} \right) \
\left(\frac{k_2^2}{\alpha}-r^2 \right) - \omega^2(1+r^2) \ .  
\ee
In order to get a closed string solution, we must assume that
\be{wcon} \Delta \ = \ 
(\alpha - \beta )^2 -\omega^2(2\alpha + 2\beta +4k_2^2 -\omega^2) > 0
\ . \ee
For $\omega^2=0$ this is automatically satisfied. However, for generic
values of $\omega$ it is not the case. We shall proceed to solve
(\ref{drh}) under this assumption, and at the end check if
(\ref{wcon}) is obeyed.
The solution for $r(\sigma)$ is similar to the one found for
$\omega=0$
\be{rsh}
r^2 (\sigma) = \frac{k_2^2/2}{\alpha \beta+k_2^2 \omega^2}
\left( (\alpha +\beta -\omega^2) 
+\sqrt{\Delta}
\sin \left[\sqrt{\frac{4\alpha \beta}{k_2^2}+4 \omega^2}
 \ \sigma \right] \right)
\ee 
For the string to close $r(\sigma)=r(\sigma+2\pi)$, we need
\be{}
n{_\omega} = \sqrt{\frac{4\alpha \beta}{k_2^2} + 4\omega^2} =
\mbox{integer} \ .\ee
The solution to the other two coordinates can be found in the same way
as for the $\omega=0$ case. In particular, the periodic boundary
conditions on the angular coordinate will lead to the same result as
in the $\omega=0$ case with $n$ replaced by $n_{\omega}$.   
The time coordinate is given by 
\be{tw}
t (\tau, \sigma) \ = \ c_1 \tau - c_2 \sigma +
\frac{(c_2+k_1)\sqrt{4\alpha \beta +4k_2^2\omega^2}}
{n_{\omega}\sqrt{(k_2^2+\alpha)(k_2^2+\beta)}}
\ee
$$\arctan \left[ \frac{\left(1+\frac{k_2^2/2}{\alpha \beta
+k_2^2\omega^2}(\alpha+\beta-\omega)\right)\tan[\frac{n_{\omega}}{2}\sigma]+
\frac{k_2^2/2}{\alpha
\beta +k_2^2\omega^2}\sqrt{\Delta}}{\sqrt{\frac{(k_2^2+\alpha)(k_2^2+\beta)}{\alpha \beta
+k_2^2 \omega^2}}}\right] 
$$
Imposing $t(\tau,0)=t(\tau,2\pi)$ gives
\be{k2w}
k_2 = \frac{2c_1^3-2c_1\omega^2}{n_{\omega}^2-4c_1^2} \ . \ee
We now return to (\ref{wcon}), and check if this assumption is
correct. We first note that from the definition of $n_{\omega}$ it 
is now bounded from below, $n_{\omega}>2\omega$. To ensure $k_2 \ge 0$
we must have $c_1 \ge \omega$. Combining all these
relations we get $n_{\omega}>2c_1 \ge 2\omega$. It can be seen by direct
substitution of $k_2$ in (\ref{wcon}) that this is enough to ensure
$\Delta \ge 0$. 
When $c_1=\omega$, $\Delta=0$. We will shortly see what is the physical
meaning of this point.

Using this result and (\ref{concha1}) we can write down the energy and spin
of the string
\be{}
E = \frac{R^2(c_1+k_2)}{\ap} \ = \ \frac{R^2 c_1}{\ap} \
\frac{n_{\omega}^2-2c_1^2-2\omega^2}{n_{\omega}^2-4c_1^2} \ ,
\ee 
$$
S \ =  \ \frac{R^2 k_1}{\ap} \ = \ \frac{R^2}{\ap} \ 
\frac{n_{\omega}(c_1^2-\omega^2)}{n_{\omega}^2-4c_1^2} \ .
$$
It is again straightforward, though lengthy, to write down
$E=E(S,n_{\omega},\omega)$. Several important limits of our result can,
however, be easily analyzed. The first 
one is by setting $\omega=0$ and retrieving the earlier 
result (\ref{concha1}).
The second one is by  taking $c_1=\omega$. 
We get a string state with vanishing spin in
$AdS_3$. At this point the energy of the string depends only on
$\omega$ in the expected way.
\be{}
E \ = \ \frac{R^2 \omega}{\ap} \ . \ee
The last interesting limit is that of ``long'' strings, which corresponds
to $\epsilon=n_{\omega}^2-4c_1^2 \ra 0$. In this limit we get  
\be{}
E \ = \ S + \frac{3}{8} \frac{R^2 n_{\omega}}{\ap} + \frac{1}{2}
\frac{R^2 \omega^2}{\ap n_{\omega}} + \cO
\left(\frac{R^4 n_{\omega}^2}{\ap^2 S} \right) \ . \ee
Again a leading linear behavior followed by a constant correction.

\section{Discussion}

In this paper we analyzed spinning strings in $AdS_3 \times S^3
\times \cM$ with NS-NS 2-form background. 
We saw that the ``planetoid''
spinning string solution of $AdS_5$ is not a solution
in $AdS_3$ because
of the NS-NS 2-form. 
 According to the AdS/CFT duality string
theory on $AdS_3 \times S^3 \times \cM$ is dual to a superconformal field
theory on a cylinder, which is the boundary of $AdS_3$ in global
coordinates.  
In fact string theory on
this background can be exactly solved in terms of the
$SL(2,R)$ WZW model and the dual (space-time) conformal field theory
can be constructed.
There is an exact
correspondence between the bulk string states and the boundary operators.   
We denote by $\cL_n$ the Virasoro
operators of the boundary superconformal theory, and by $\cJ^3$ the
generator of $U(1) \subset  SU(2)$ subgroup of the $(4,4)$ superconformal
algebra. These are not to be confused with $L_0$ and $J^3$ of the bulk
WZW model. The relation to the bulk parameters $E$, $S$ and $\omega$ is
\be{}
E = \cL_0 + \bar{\cL}_0 \ , \qquad S = \cL_0 - \bar{\cL}_0\ ,
\qquad \cJ^3 = \cJ^3 =\omega \ . \ee
Using the known relation between the space-times energy in $AdS_3$ and
the eigenvalues of $J_0^3$ and $\bar{J}_0^3$ from the WZW model we get
that $\cL_0=J_0^3$ and $\bar{\cL}_0=\bar{J}_0^3$. 
Thus, for a string state with a given $(m,\bar{m})$ (\ref{mmbar})
we know the exact space-time energy $E=m+\bar{m}$ and
spin $S=m-\bar{m}$. The classical string solutions that we found also give
the information about the general semi-classical relation between
the two, $E(S)$.
We saw that the leading
behavior of the relation is the same as in $AdS_5$, but the anomalous correction is
constant rather than logarithmic.

In comparison the S-dual background has R-R charge.
The conformal theory in this case can be thought of the IR fixed point
of a $(1+1)$-dimensional gauge theory on the D1-D5 system. A spinning
fundamental string in this background is of the same, ``planetoid'',  
form as the GKP string, and is likely to be dual to a similar operator
on the conformal theory side. 
In this case the relation between the energy and spin of the
string exhibits the same $\ln S$ behavior in the ``long'' string limit.
We saw that 
the conformal field theory dual to the  $AdS_3$ NS-NS background  is quite
different in this respect. 

\vskip 2cm

{\bf Acknowledgments.}
We would like to thank A. Giveon and M. Karliner
for a valuable discussion.
The work of AL was supported in part by NSF grant-0071512.
The work of YO is supported in part by 
the US-Israel Binational Science Foundation.

\vskip 2cm

\end{document}